\documentclass[reprint,amsmath,amssymb,aps,prl,superscriptaddress]{revtex4-1}

\usepackage{graphicx}
\usepackage{dcolumn}
\usepackage{bm}
\usepackage{hyperref}

\usepackage{xcolor}

\usepackage{subcaption}
\newcommand{\labelphantom}[1]{%
{\phantomsubcaption%
\label{#1}}%
}%


\begin{document}

\title{Robust avoidance of edge-localized modes alongside gradient formation in the negative triangularity tokamak edge}

\author{A. O. Nelson}
\affiliation{Columbia University, New York, NY 10027, USA}

\author{L. Schmitz}
\affiliation{University of California – Los Angeles, Los Angeles, CA 90095, USA}

\author{C. Paz-Soldan}
\affiliation{Columbia University, New York, NY 10027, USA}

\author{K. E. Thome}
\affiliation{General Atomics, San Diego, California 92186, USA}

\author{T. B. Cote}
\affiliation{General Atomics, San Diego, California 92186, USA}

\author{N. Leuthold}
\affiliation{Columbia University, New York, NY 10027, USA}

\author{F. Scotti}
\affiliation{Lawrence Livermore National Laboratory, Livermore, CA, USA}

\author{M. E. Austin}
\affiliation{The University of Texas at Austin, Austin, Texas 78712, USA}

\author{A. Hyatt}
\affiliation{General Atomics, San Diego, California 92186, USA}

\author{T. Osborne}
\affiliation{General Atomics, San Diego, California 92186, USA}

\date{\today}

\begin{abstract}

In a series of 
high performance diverted discharges on DIII-D, we demonstrate that strong 
negative triangularity (NT) shaping robustly 
suppresses all edge-localized mode (ELM) activity over a wide range of plasma conditions: $\langle n\rangle=0.1-1.5\times10^{20}$\,m$^{-3}$, $P_\mathrm{aux}=0-15$\,MW and $|B_\mathrm{t}|=1-2.2$\,T, corresponding to $P_\mathrm{loss}/P_\mathrm{LH08}\sim8$. 
The full dataset is consistent with the theoretical prediction that magnetic shear in the NT edge inhibits access to ELMing H-mode regimes; all experimental pressure profiles are found to be at or below the infinite-$n$ ballooning stability limit.
Importantly, we also report enhanced edge pressure gradients at strong NT that are significantly steeper than in traditional ELM-free L-mode plasmas 
and provide significant promise for NT reactor integration. 

\end{abstract}

\maketitle


Magnetic fusion energy reactors must achieve significant plasma pressures alongside sufficiently high energy confinement times in order to achieve the high fusion gain ($Q\sim10$) needed for energy production. The predominant approach employed to meet these requirements involves tokamak operation in a high-confinement (H-mode) scenario with positive triangularity (PT), which features a region of steep pressure gradients near the plasma edge called the pedestal \cite{Wagner1984}. While strong pedestals raise the core plasma pressure, they also beget violent instabilities called edge-localized modes (ELMs) that periodically connect the hot core plasma to the cooler edge region and deposit tremendous heat fluxes on the machine wall \cite{zohm_edge_1996, Leonard2014}. In a reactor, it is expected that ELMs will be powerful enough to cause significant and potentially fatal damage to plasma-facing components \cite{Gunn2017}, necessitating development of a reactor scenario that operates at high performance while simultaneously remaining completely ELM-free \cite{paz-soldan_plasma_2021}. 

Over the past few decades, numerous ELM avoidance strategies have been pursued as potential solutions to this power-handling problem. These include quiescent H-modes (QH-mode) \cite{Burrell2001}, ELM suppression with resonant magnetic perturbations (RMPs) \cite{suttrop_experimental_2018}, improved-confinement (I-mode) scenarios \cite{Whyte2010}, highly radiative low-confinement (L-mode) scenarios \cite{frank_radiative_2022} and enhanced $D_\mathrm{alpha}$ H-modes \cite{Greenwald1999}, among others. Each of these techniques achieves ELM suppression through the introduction of an additional transport-inducing process in the plasma edge that prevents access to standard ELMing H-mode operation, often at some (manageable) expense in overall plasma performance. While it is hoped that some of these techniques will be applicable during high performance plasma operation on machines like ITER, they are each subject to different access criteria that are difficult to robustly extrapolate towards reactors \cite{paz-soldan_plasma_2021, viezzer_prospects_2023}. 

Recently, extensive work on the TCV \cite{pochelon_energy_1999, Camenen2007}, DIII-D \cite{Austin2019, marinoni_diverted_2021} and AUG \cite{happel_overview_2022} tokamaks has renewed interest in an additional ELM-avoidance strategy involving operation with negative triangularly (NT) shaping. The plasma triangularity ($\delta$) is defined as the average of the upper and lower triangularities $\delta_\mathrm{u,l} \equiv (R_\mathrm{geo}-R_\mathrm{u,l})/a_\mathrm{minor}$, where $R_\mathrm{geo}$ is the geometric major radius, $R_\mathrm{u,l}$ are respectively the major radius of the highest and lowest points along the plasma separatrix, and $a_\mathrm{minor}$ is the minor radius of the plasma. Early experiments in this regime \cite{pochelon_energy_1999, Camenen2007, Austin2019, marinoni_diverted_2021} have been able to achieve strong core performance (confinement factors \cite{ITERPhysicsBasisEditors1999} of $H_\mathrm{98y2}>1$, and normalized betas of $\beta_\mathrm{N}>2.5$) but have been unable to access ELMing H-mode beyond a critical triangularity $\delta<\delta_\mathrm{crit}$, optimistically indicating that NT operation could offer an additional solution to the power-handling problem faced by tokamak plasmas \cite{Medvedev2015, Kikuchi2019}. 

In this Letter, we present new data from a high-power NT campaign on the DIII-D tokamak to demonstrate that robust ELM avoidance is a fundamental property of the NT edge and that it does not inhibit access to high core pressure. A unique set of carbon plasma facing components is installed in the DIII-D tokamak that allows for strongly-shaped, diverted NT operation at record heating powers and densities \cite{nelson_vertical_2023}, showing that NT robustly prevents ELM instabilities over an all-encompassing parameter space as long as $\delta<\delta_\mathrm{crit}$. A particular class of magnetohydrodynamic (MHD) instability, the ideal ballooning mode, is identified as a fundamental gradient-limiting mechanism in the NT edge that allows for pedestal formation while avoiding the instability limits responsible for triggering ELMs, confirming theoretical predictions from previous work \cite{nelson_prospects_2022}. 
With the novel analysis presented here, we find that this edge state is fundamentally different than standard L-mode plasmas in positive triangularity. As such, we propose that it is more accurately described as an ``NT edge" than ``L-mode.'' 


The leading understanding of H-mode behavior identifies coupled, finite-$n$ peeling-ballooning (PB) modes as the fundamental MHD instability responsible for triggering ELMs in the plasma edge \cite{Leonard2014}. Peeling-ballooning instabilities are destabilized both by strong edge current ($j_\mathrm{edge}$) and by strong normalized pressure gradients ($\alpha$), thereby setting a hard upper limit on the conditions achievable in the plasma edge. Here $\alpha$ is defined as
\begin{equation}
    \label{eq:alpha}
    \alpha = \frac{\mu_\mathrm{0}}{2\pi^2}\frac{\partial V}{\partial\psi}\bigg(\frac{V}{2\pi^2R_\mathrm{0}}\bigg)^{1/2}\frac{dp}{d\psi},
\end{equation}
where $V$ the volume enclosed by each flux surface, $\psi$ the poloidal flux, $p$ the plasma pressure and $R_\mathrm{0}$ the plasma major radius. To avoid triggering ELMs, additional edge transport must be induced in order to ensure that $j_\mathrm{edge}$ and $\alpha$ are held some distance below the PB limit. This is achieved, for example, through the edge harmonic oscillation (EHO) in QH-mode plasmas \cite{Burrell2001} and through the weakly coherent mode (WCM) in I-mode \cite{Whyte2010}. 

\begin{figure}
    \includegraphics[width=1\linewidth]{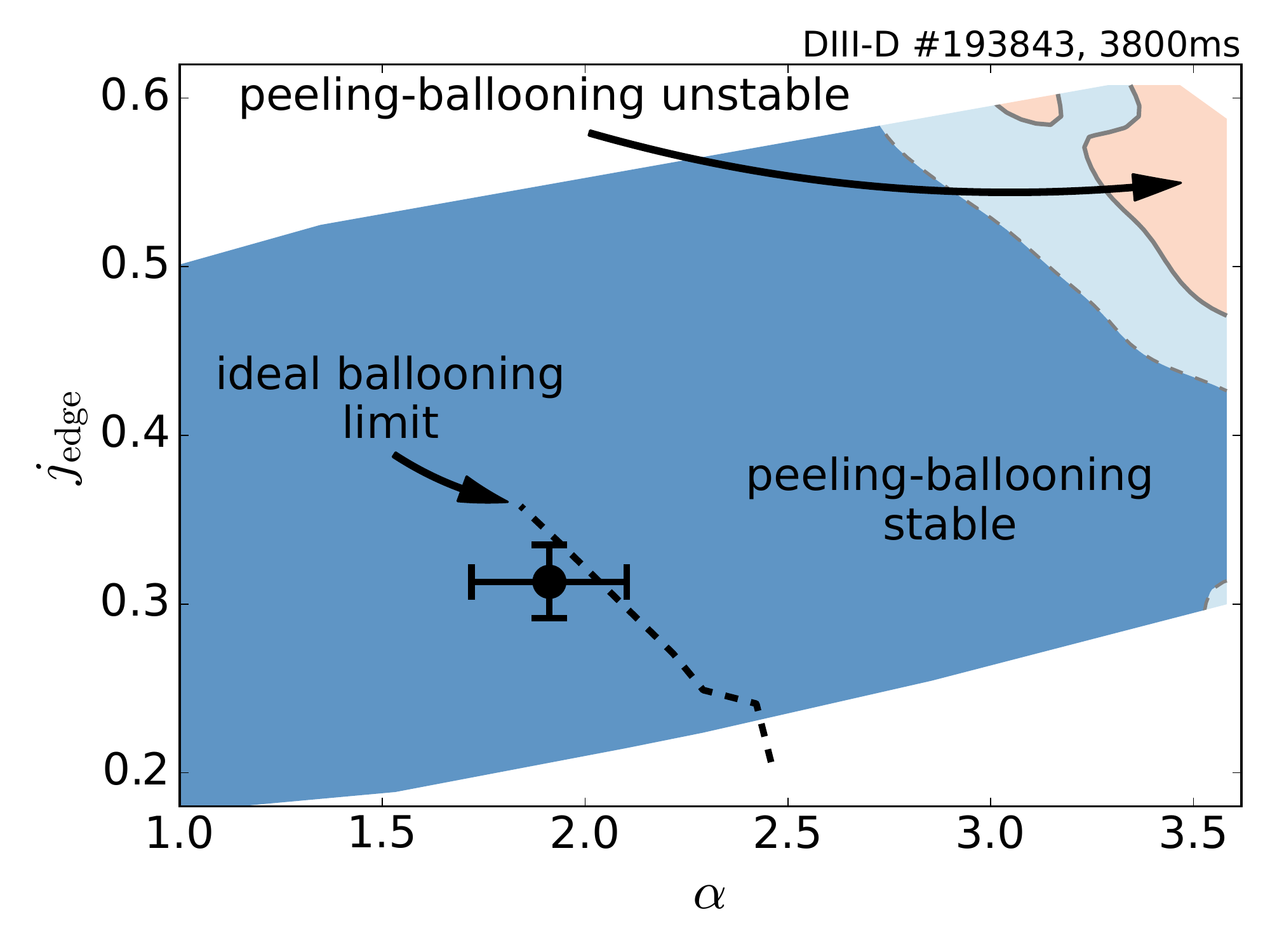}
    \caption{Calculation of the infinite-$n$ ballooning (dashed line) and peeling-ballooning (blue-red) limits for a typical high-performance NT plasma on DIII-D. The measured experimental point lies directly at the ideal ballooning limit, safely below the ELM instability boundary.}
    \label{fig:ELITE}
\end{figure}

In NT plasmas, the ideal (infinite-$n$) ballooning mode has been proposed as the ultimate gradient-limiting mode responsible for ensuring that the PB instability is not reached \cite{saarelma_ballooning_2021, nelson_prospects_2022}. As described in reference~\cite{nelson_prospects_2022}, the local magnetic shear $s$ 
has a peak near the separatrix x-points for tokamak plasmas and a minimum on the outboard midplane. Notably, the shear stabilization for ballooning modes is proportional to $s^2$ \cite{bishop_stability_1986}. In NT, the x-points are radially farther from the machine center than the equilibrium magnetic axis, localizing the maximum in $s$ to the destabilizing ``bad curvature'' region of the plasma. This means that NT geometries force a null-crossing in $s$ to appear in the bad curvature edge region, which destabilizes ideal ballooning modes at the so-called ``1$^{st}$ stability limit" \cite{nelson_prospects_2022}. In conventional PT geometries, these modes are stabilized throughout the entire bad curvature region, opening a window to a 2$^{nd}$ stability region that supports gradient growth typical of an H-mode pedestal. For a typical high-performance NT plasma on DIII-D with $\delta\sim-0.5$, the ideal ballooning and peeling-ballooning limits are calculated using the MHD stability codes BALOO \cite{miller_stable_1997} and ELITE \cite{Snyder2002}, respectively, and plotted in figure~\ref{fig:ELITE}. Notably, the ideal ballooning limit appears at a lower $\alpha$ than the PB limit, preventing the triggering of ELMs by limiting edge gradient growth in manner akin to the QH-mode EHO or I-mode WCM. 

\begin{figure}
    \includegraphics[width=1\linewidth]{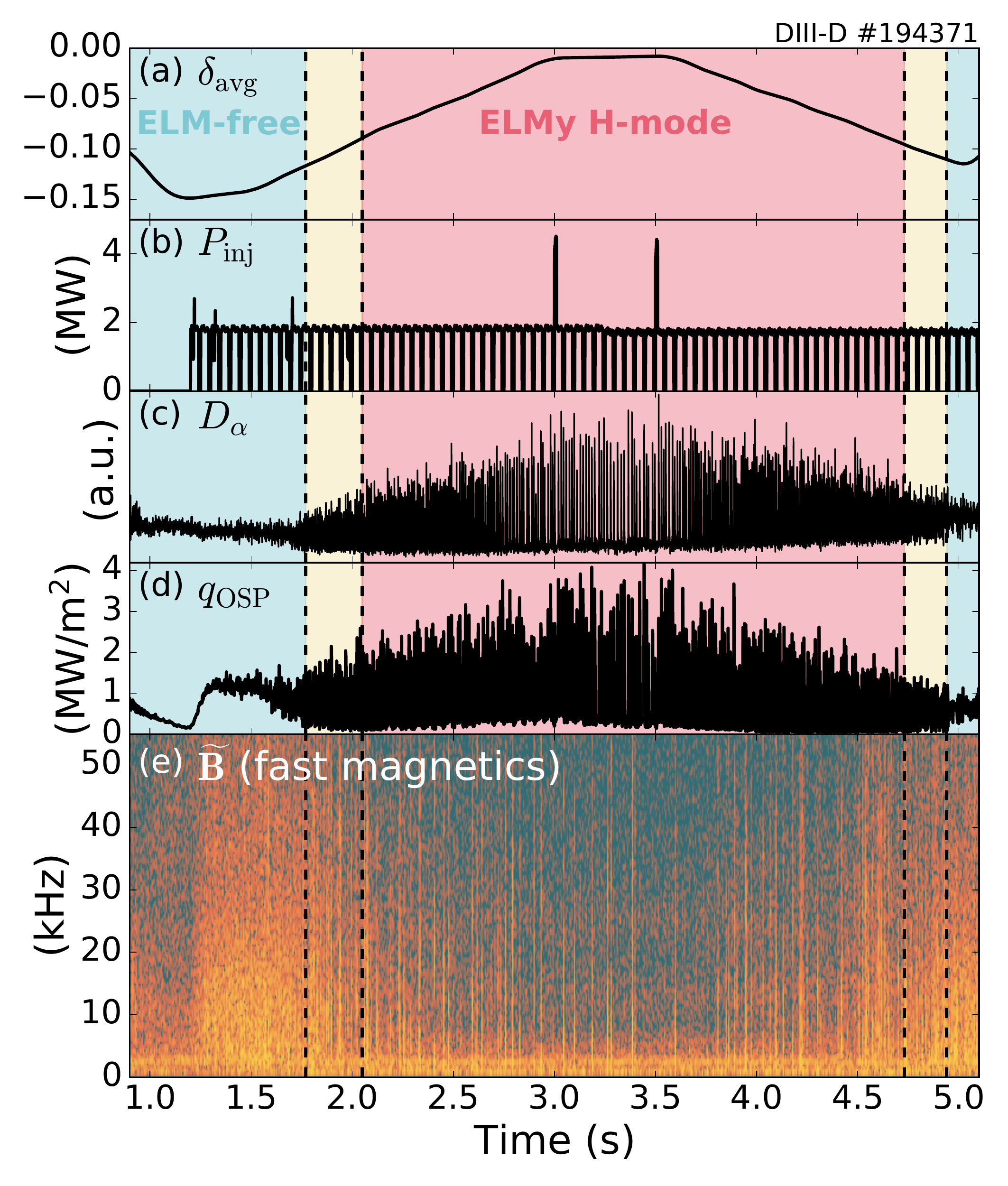}
    \labelphantom{fig:time-a}
    \labelphantom{fig:time-b}
    \labelphantom{fig:time-c}
    \labelphantom{fig:time-d}
    \labelphantom{fig:time-e}
    \caption{As the triangularity (a) is varied at constant input power (b), the plasma transitions smoothly from an ELM-free state (blue) to an ELMing H-mode regime (red) through a oscillatory transition (yellow). Measurements of the $D_\mathrm{\alpha}$ line emission (c) and divertor heat flux (d) demonstrate that peak power incident on the machine walls is strongly reduced during operation at strong NT. (e) Inspection of high-frequency magnetic fluctuations during this time reveals enhanced turbulent activity during the ELM-suppressed periods.}
    \label{fig:time}
\end{figure}

To demonstrate the sensitivity of NT ELM suppression on the plasma shape, in figure~\ref{fig:time} the triangularity $\delta$ is varied in time at constant heating power ($P_\mathrm{inj}\approx2$\,MW)
. Small changes in $\delta$ near $\delta_\mathrm{crit}\sim-0.12$ prompt smooth transitions from an ELM-free regime to an ELMy H-mode. In contrast to the expected behavior in PT plasmas, the transition between the ELM-free state and the ELMy H-mode state is not realized experimentally as an abrupt phase transition, but rather evolves smoothly through a transitional dithery phase as the stability window is slowly widened. This is evidenced by the slowly evolving $D_\mathrm{alpha}$ emission measurement shown in figure~\ref{fig:time-c}. Figure~\ref{fig:time-d} demonstrates a reduction in the peak outer strikepoint heat flux ($q_\mathrm{OSP}$) during the periods when ELMs are suppressed, highlighting the potential for NT configurations to ameliorate the power-handling problem presented by ELMs. Fourier analysis of high-frequency magnetic measurements shows strong broadband fluctuations in the ELM-free state after heating power is introduced (figure~\ref{fig:time-e}), consistent with enhanced turbulence when the ballooning stability limit is reached. Notably, no strong hysteresis effect is observed when entering or leaving the ELM-free state, again consistent with physics dominated by ideal MHD activity.


\begin{figure}
    \includegraphics[width=1\linewidth]{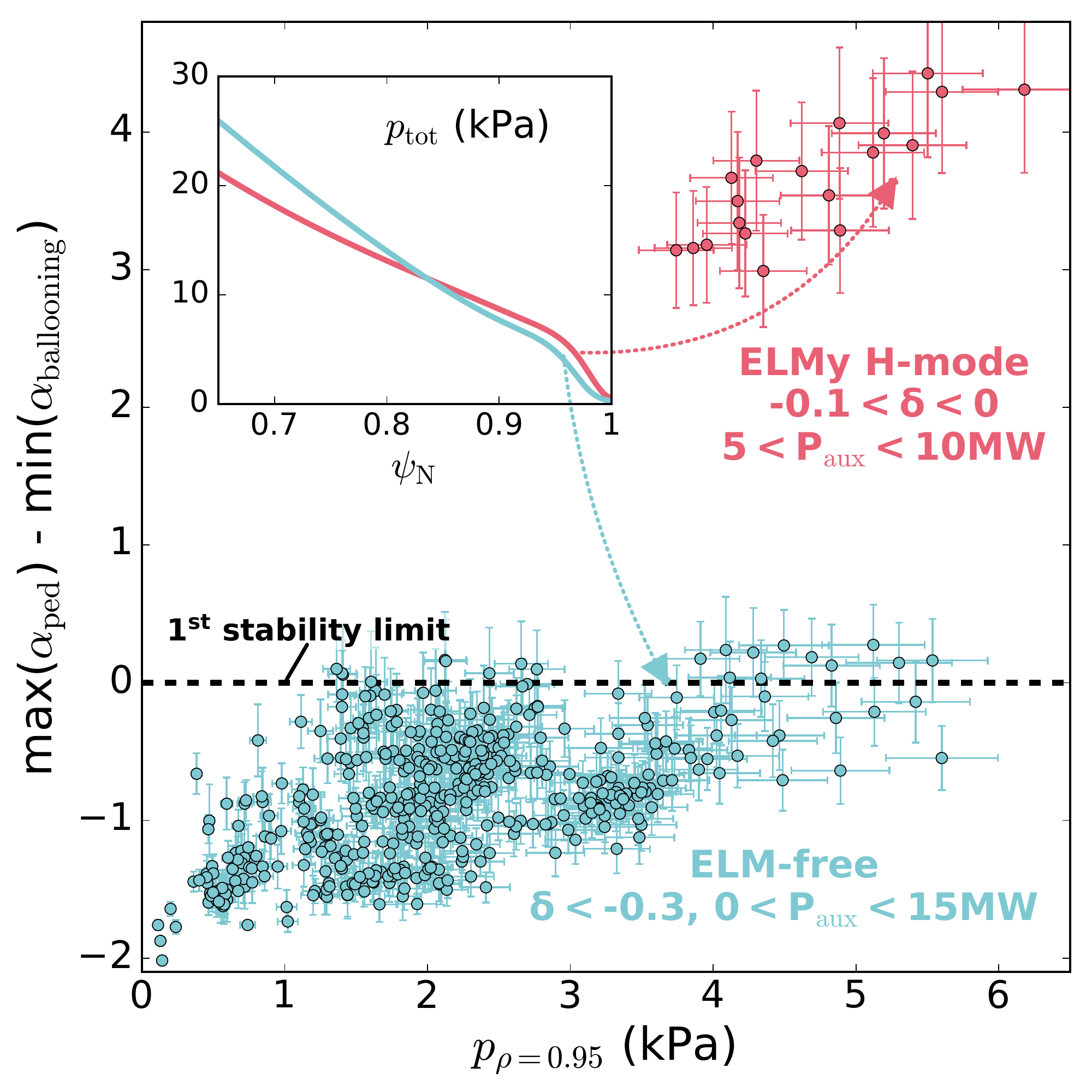}
    \caption{Within experimental error bars, all DIII-D discharges with strong NT ($\delta<-0.3$, blue) are limited by the 1$^{st}$ ballooning stability limit. For comparison, a selection of ELMy H-mode discharges with weaker $\delta$ and access to the 2$^{nd}$ stability region are shown in red. Inset: pressure profiles for representative ELM-free and ELMy H-mode cases, both with $P_\mathrm{aux}=8\,$MW.}
    \label{fig:balldb}
\end{figure}

Unlike the gradient-limiting mechanisms responsible for ELM suppression in other regimes, destabilization of the ideal ballooning mode is a direct consequence of the magnetic geometry and is thus entirely robust to changes in plasma conditions. To demonstrate the simplicity of this effect, we utilize recent developments in automatic kinetic equilibrium reconstructions with the CAKE code \cite{xing_cake_2021} to visualize ideal ballooning stability on a database level. Equilibrium reconstructions for 425 ELM-free time slices from equilibria with $\delta<-0.3$ are analyzed and presented in figure~\ref{fig:balldb}. This dataset covers a wide range of plasma conditions with auxiliary heating powers up to $P_\mathrm{aux}\sim15\,$MW, volume averaged densities of $\langle n\rangle=0.1-1.5\times10^{20}$\,m$^{-3}$, applied torques of $T_\mathrm{inj}=-4$ to $10$\,N-m, plasma currents of $0.3<I_\mathrm{p}<1.1\,$MA and on-axis magnetic fields of $1<|B_\mathrm{t}|<2.2\,$T. 

In figure~\ref{fig:balldb}, the distance between the equilibrium $\alpha$ and the minimum $\alpha$ required for ballooning mode destabilization is plotted as a function of the edge pressure $p_\mathrm{\rho=0.95}$. The strong NT discharges cover a range of edge stability, but no ELM-free discharges support larger gradients than allowed by ideal ballooning stability. This behavior shows that, while particular discharges in this dataset may be limited by additional physics mechanisms (the subject of future reports), the infinite-$n$ ballooning mode sets an upper limit on the pressure gradient in the NT edge. For comparison, a selection of strong ELMy H-mode discharges are also shown in figure~\ref{fig:balldb}. In these cases, which require weak $\delta$, the plasma is able to access the 2$^{nd}$ stability region through additional growth of the density pedestal. The ELMy H-mode plasmas therefore achieve normalized gradients well above the 1$^{st}$ stability limit that dominates behavior at strong NT. The inset of figure~\ref{fig:balldb} shows a comparison of the pressure profiles for representative ELM-free and ELMy H-mode NT shots. As expected, the profile gradient in the edge region is much steeper in the weak NT H-mode case. However, it is also immediately evident that the diminished edge pressure gradient of the ELM-free case does not limit total plasma performance, as the ELM-free NT edge enables higher pressure gradients throughout the core region. 

\begin{figure}
    \includegraphics[width=1\linewidth]{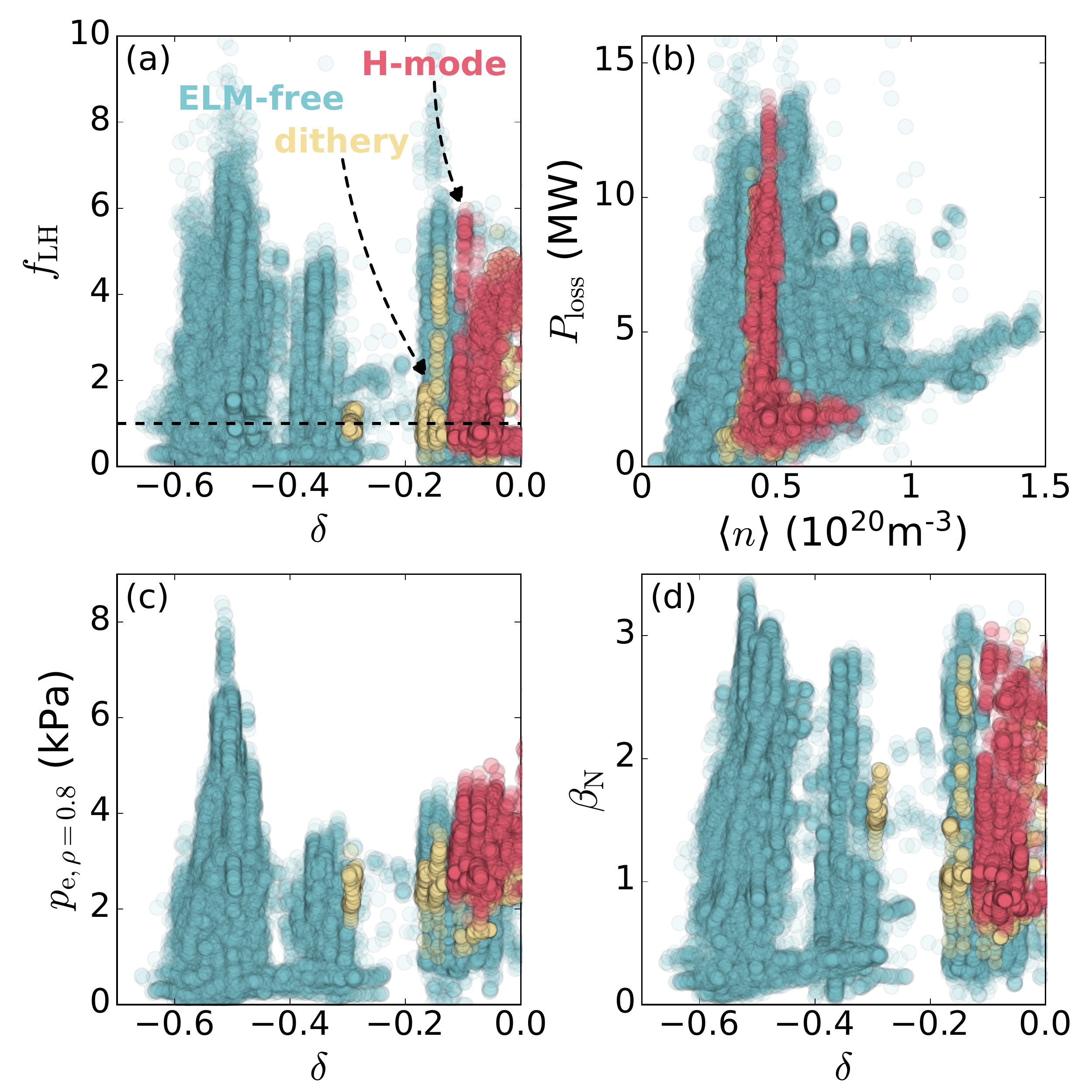}
    \labelphantom{fig:db-a}
    \labelphantom{fig:db-b}
    \labelphantom{fig:db-c}
    \labelphantom{fig:db-d}
    \caption{(a) For the entire NT database on DIII-D, the H-mode threshold power fraction $f_\mathrm{LH}$ is shown as a function of $\delta<0$. ELMy H-modes, which only occur at weak $\delta$ are colored in red, oscillatory periods in yellow, and ELM-free periods in blue. (b) The volume averaged density $\langle n \rangle$ and separatrix power $P_\mathrm{loss}$ reveal the breadth of the ELM-free space. (c) The edge pressure (at radius $\rho=0.8$) and (d) $\beta_\mathrm{n}$ are shown as a function of $\delta<0$ to demonstrate enhanced performance achievable with the ELM-free NT edge.}
    \label{fig:db}
\end{figure}

In figure~\ref{fig:db}, we further characterize the edge behavior by examining time slices every 20\,ms throughout 850 separate discharges (the entire DIII-D NT database). Figure~\ref{fig:db-a} shows access to the NT ELM-free state as a function $\delta$ and H-mode threshold power fraction $f_\mathrm{LH}=P_\mathrm{loss}/P_\mathrm{LH08}$, where 
\begin{equation}
    P_\mathrm{loss} = P_\mathrm{aux} + P_\mathrm{Ohmic} - P_\mathrm{rad, core} - \frac{dW_\mathrm{MHD}}{dt}
    \label{eq:ploss}
\end{equation}
is a measure of the power crossing the separatrix and
\begin{equation}
    P_\mathrm{LH08} = 2.15 \, \langle n\rangle^{0.782} B_\mathrm{t}^{0.772} a_\mathrm{minor}^{0.975} R_\mathrm{0}^{0.999}
    \label{eq:Martin}
\end{equation}
is the typical threshold power needed for H-mode access based on the scalings in reference~\cite{Martin2008}. (Here the line-averaged plasma density $\langle n \rangle$ is given in [$10^{20}$ m$^{-3}$], $B_\mathrm{t}$ in [T], and the radii $a_\mathrm{minor}$ and $R_\mathrm{0}$ in [m].) ELM-free operation is regularly achieved on DIII-D below a critical triangularity of $\delta_\mathrm{crit}\sim-0.15$, and that dithery regimes (from figure~\ref{fig:time}) only exist at $\delta\gtrsim-0.3$, even when heating the plasma with upwards of eight times the expected H-mode threshold power $P_\mathrm{LH08}$. In figure~\ref{fig:db-b}, we show a traditional view of the L-to-H transition space expected for PT discharges by comparing the volume average density to the total power crossing the separatrix ($P_\mathrm{loss}$). It is readily observed that the ELM-free operating space occupies a broader parameter space than NT ELMy H-modes, highlighting that ELM-suppression in NT is insensitive to control parameters and thus fundamentally different than in PT. 


In addition to demonstrating robust ELM suppression across the entire operational space, figure~\ref{fig:db-c} shows how the edge pressure $p_\mathrm{e}$ (measured at a normalized radius of $\rho=0.8$) is impacted by the ELM-free NT edge. H-mode access predictably leads to enhanced $p_\mathrm{e}$ at small $|\delta|$, before the ideal ballooning physics takes effect. However, the lack of H-mode at stronger $\delta$ does not prevent the ELM-free scenarios from reaching or exceeding edge pressures achieved by H-modes in weaker shapes. 
Normalized measures of global performance tell a similar story over the DIII-D NT operational space, as shown by in figure~\ref{fig:db-d}. Neither $\beta_\mathrm{N}$ nor $H_\mathrm{98y2}$ is hindered by the ELM suppression mechanisms of NT. Instead, a large portion of the DIII-D NT database meets reasonable reactor targets of $H_\mathrm{98y2}>1$, $\beta_\mathrm{N}>2.5$ and $f_\mathrm{GW}>1$ at triangularities of $\delta\sim-0.5$, well within the ELM-free operational space. Indeed, the highest preforming NT discharges achieved to date on DIII-D are completely ELM-free.

\begin{figure}
    \includegraphics[width=0.8\linewidth]{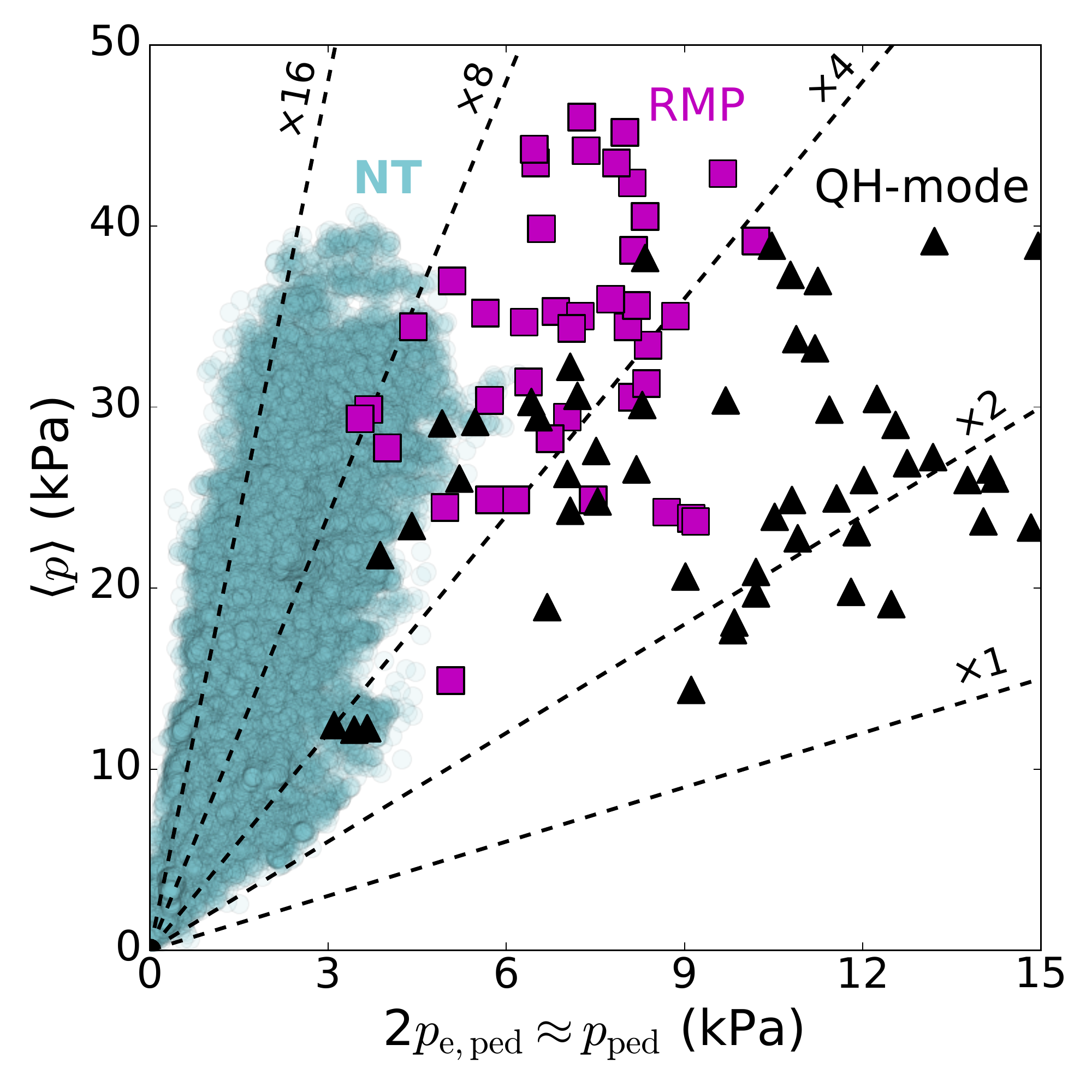}
    \caption{The volume averaged pressure $\langle p \rangle$ and pedestal pressure $p_\mathrm{ped}$ for ELM-free NT discharges (blue), PT RMP ELM-suppressed H-modes (magenta) and PT QH-modes (black) on DIII-D are compared.}
    \label{fig:RMPQH}
\end{figure}

The high performance achieved by the ELM-free NT edge is also demonstrated in figure~\ref{fig:RMPQH}, which compares the volume averaged pressures $\langle p \rangle$ and edge pressures $p_\mathrm{ped}\approx2\times p_\mathrm{e,ped}$ for the ELM-free NT database and the DIII-D RMP ELM-suppressed and QH-mode databases (both in PT, from reference~\cite{paz-soldan_plasma_2021}). As expected, both the RMP and QH-mode discharges feature significantly larger pedestal pressures than the NT configurations due to their H-mode characteristics. However, the maximum achievable $\langle p \rangle$ between all three regimes remains quite comparable, highlighting that the ELM-free NT edge does not inhibit steep profile gradients from forming in the core region where fusion power will be concentrated. We note here that many of the RMP and QH mode discharges included in figure~\ref{fig:RMPQH} access an ELM-free regime for only a portion ($\lesssim30$\%) of the full shot, while NT discharges included are entirely ELM free for the full discharge duration. This extremely robust nature of NT ELM avoidance is a unique and promising feature among the various ELM suppression techniques currently under investigation worldwide.

Finally, we would like to reiterate that the traditional relationship between L-mode and H-mode as established by decades of experience in PT does not apply to discharges with strong NT. Negative triangularity plasmas are \textit{not} kept out of H-mode via a lack of power crossing the separatrix, as is typically assumed of L-mode plasmas in PT: no evidence of an LH transition threshold power in plasmas with strong enough NT shaping has yet been encountered. Further, ELM-free scenarios in NT can have profile gradients similar to those observed in NT ELMy H-mode scenarios. These plasmas occupy a unique space in the operational domain for tokamaks (naturally ELM-free operation at high normalized performance) and are held there regardless of plasma conditions due to a unique physical mechanism (gradient limiting via ideal ballooning modes in the plasma edge). As such, we refer to these plasmas not by ``L-mode'' as suggested by the current literature, but rather propose that they be categorized simply as having an ELM-free NT edge.



\begin{acknowledgments}
\textit{Acknowledgments.} Part of data analysis for this work was performed using the OMFIT integrated modeling framework \cite{Meneghini2015, Logan2018}. This material was supported by the U.S. Department of Energy, Office of Science, Office of Fusion Energy Sciences, using the DIII-D National Fusion Facility, a DOE Office of Science user facility, under Awards DE-SC0022270, DE-SC0022272, DE-SC0020287, DE-FG02-97ER54415 and DE-FC02-04ER54698. This report is prepared as an account of work sponsored by an agency of the United States Government. Neither the United States Government nor any agency thereof, nor any of their employees, makes any warranty, express or implied, or assumes any legal liability or responsibility for the accuracy, completeness, or usefulness of any information, apparatus, product, or process disclosed, or represents that its use would not infringe privately owned rights. Reference herein to any specific commercial product, process, or service by trade name, trademark, manufacturer, or otherwise, does not necessarily constitute or imply its endorsement, recommendation, or favoring by the United States Government or any agency thereof. The views and opinions of authors expressed herein do not necessarily state or reflect those of the United States Government or any agency thereof.
\end{acknowledgments}



%

\end{document}